\documentclass[nohyper,12pt,letterpaper]{JHEP}
\usepackage{epsfig}

\newfont{\frak}{eufm10 scaled 1200}

\newfont{\Bbb}{msbm10 scaled 1200}     
\newcommand{\mathbb}[1]{\mbox{\Bbb #1}}
\DeclareSymbolFont{AMSa}{U}{msa}{m}{n}
\DeclareSymbolFont{AMSb}{U}{msb}{m}{n}
\let\Box\relax
\DeclareMathSymbol{\Box}{\mathord}{AMSa}{"03}

\def \eqn#1#2{\begin{equation}#2\label{#1}\end{equation}}

\title{Cosmological Breaking of Supersymmetry\\
         {\it or}\\
       Little Lambda Goes Back to the Future II}

\author{T. Banks\thanks{On Leave from Rutgers University}\\
  Department of Physics and Institute for Particle Physics\\
  University of California, Santa Cruz, CA 95064\\
E-mail: \email{banks@scipp.ucsc.edu}}

\abstract{It is conjectured that M-theory in asymptotically 
flat spacetime must be supersymmetric, and that the observed
SUSY breaking in the low energy world must be attributed to
the existence of a nonzero cosmological constant.  This 
would be consistent with experiment, if the 
{\it critical exponent} $\alpha$ 
in the relation $M_{SUSY} \sim M_P (\Lambda /M_P^4)^{\alpha}$
took on the value $1/8$ , rather than its classical value 
$1/4$.  We attribute this large renormalization 
to the effect of large virtual black holes via the UV/IR correspondence.}

 
\received{???????? ?st, 2000} \accepted{???????? ?th, 1998}
 
\preprint{\hepth{0007146}\\RUNHETC-2000-24; SCIPP-23/00}
\begin{document}

 
 \section{Introduction}

This paper is an expanded version of a short talk I gave at 
Lenny
Susskind's 60th birthday celebration at Stanford University.  It
is dedicated to Lenny, who taught me how to think about physics,
and whose own recent ideas have profoundly influenced those I am
reporting on here. The central message of the paper can be
summarized in a few sentences: {\it The Bekenstein-Hawking Entropy
of (Asymptotically) DeSitter (AsDS) spaces represents the
logarithm of the total number of quantum states necessary to
describe such a universe.  This implies that the cosmological
constant is an input to the theory, rather than a quantity to be
calculated.  The structure of an AsDS universe automatically
breaks supersymmetry (SUSY). From this point of view, the \lq\lq
cosmological constant problem\rq\rq is the problem of explaining
why the SUSY breaking scale is so much larger than that associated
in classical supergravity (SUGRA) with the observed value of (bound
on ?) the cosmological constant. I suggest that large
renormalizations of the classical formula are to be expected on
the basis of the UV/IR correspondence in M theory. These may be
viewed as contributions from virtual black holes. The
phenomenologically correct formula $M_{SUSY} \sim (\Lambda M_P^4
)^{1/8} $ may be derivable from such considerations.}

The implication of these ideas is that SUSY breaking vanishes in
the flat space limit, which is consistent with the fact that we
have not succeeded in finding a string vacuum with broken SUSY and
asymptotically flat spacetime.  We will begin the paper with a
brief review of the evidence for this.

We then turn to a defense of the contention that an asymptotically
DeSitter (AsDS) universe can be described by a finite number of
states, given by the Bekenstein-Hawking formula.  We discuss the
difference between such an AsDS universe and the temporary
DeSitter phase of an inflationary universe.  We caution the reader
that once he accepts these arguments, he will be forced to
conclude that the cosmological constant is an input, or boundary
condition, rather than a parameter to be calculated.  The
conventional cosmological constant problem can then be rephrased
as: why isn't the scale of SUSY breaking related to the
cosmological constant by the standard classical SUGRA formula,
which (without fine tuning) predicts $M_{SUSY} \sim
\Lambda^{1/4}$.

We argue that this formula may receive large renormalizations.
Indeed, standard field theory calculations predict logarithmically
divergent renormalizations (at finite orders in the loop
expansion) of the masses of particles in softly broken SUSY
theories.  Conventionally, these divergences are absorbed into the
parameters in the low energy effective field theory.  Wilsonian
renormalization group arguments suggest that these divergent terms
have no dependence on the cosmological constant if this parameter
is much smaller than the cutoff scale.  We argue that the UV/IR
correspondence of M theory suggests a possible source for such a
dependence.  The highest energy states of the theory are huge
black holes with a size of order the DS horizon size.  Their
spectrum is very sensitive to the value of the cosmological
constant.  We speculate that it may change the value of the
``critical exponent'' relating the SUSY breaking scale to the
cosmological constant, to $M_{SUSY} \sim (\Lambda M_P^4 )^{1/8}$.
This formula fits the observational data.

\section{Vacuum selection }

One of the unfortunate features of M-theory as a theory of the
real world, is its plethora of unphysical, exactly SUSic vacua.
On the one hand, this aspect of the theory is precisely what has
enabled us to get so much mathematical control over its properties. On
the other hand, even if one succeeded in finding a SUSY breaking
vacuum which precisely describes the real world (we should be so
lucky!) one would still have the uncomfortable task of explaining
why the universe does not resemble one of the beautifully SUSic
vacua.

From this point of view, it is interesting and exciting that it
appears very difficult to break SUSY in a way which leaves us with
an approximately flat spacetime.   Many candidate SUSY violating
vacuum states of M-theory have tachyonic instabilities.  Almost
all\footnote{The exceptions are the models of \cite{ks}. These
have no potential for moduli at one and two loops.  I am
suspicious of these models because they appear to have an infinite
fermi-bose degeneracy, despite the absence of SUSY.  I suspect
that this degeneracy is lifted at some order of perturbation
theory, at which point a potential is generated.} classical
candidate vacua generate potentials for their moduli in the
quantum theory.  The effect of these potentials is either to drive
the system into a region of moduli space where we are unable to
analyze it (except to conclude that, since it must have a large
negative vacuum energy, it cannot describe an asymptotically flat
spacetime), or to drive it deep into the weak coupling regime
where gravity becomes a free field theory.  In neither case do we
get an acceptable description of the real world.  The weakly
coupled system has massless moduli and low energy effective
parameters which vary too rapidly with time.

The above analysis is based on weakly coupled string theory and
semiclassical SUGRA.  Similar conclusions follow from an analysis
of SUSY breaking in Matrix Theory\cite{bfssplus}\cite{bmotl} .  In
this nonperturbative formulation of M-theory in a variety of
asymptotically flat, SUSY, spacetimes, asymptotic spacetime (more
precisely the configuration space of multiparticle asymptotic
states propagating on the spacetime) arises as the moduli space of
a SUSY quantum system.  Breaking SUSY collapses spacetime.

By analogy with the AdS/CFT correspondence for asymptotically AdS
SUSY vacua, one might try to find a nonsupersymmetric version of
this correspondence (with an AdS space with curvature much less
than the Planck scale) by searching for conformal field theories
with certain properties. In particular, they should have a large
gap in dimensions between the stress tensor, and all but a small
number of other operators in the theory.  The stress tensor is the
primary field corresponding to gravitons in AdS, while other
operators correspond to states with mass of order the string or
Planck scale.  The gap in dimensions indicates a large ratio
between the AdS curvature scale and the Planck scale.  In SUSY
examples this gap is "guaranteed" by SUSY nonrenormalization
theorems combined with an {\it hypothetical} scaling law for the
dimensions of nonchiral operators in the large $g_s N$ limit. The
dimension of the stress tensor is always protected, but in the
absence of SUSY we do not expect to have lines of fixed points and
there is no obvious parameter which could tune the gap to be
asymptotically large.

Another feature of a large AdS space which would have to be 
reproduced by our hypothetical conformal field theory, 
would be the existence of multigraviton excitations.  Even in 
supersymmetric examples this property is not well understood.
That is, it is understood only in the regime of $AdS_5 \times S^5$
moduli space where the 't Hooft expansion is applicable.
And in this regime, multiparticle excitations exist 
even when the curvature of AdS space is large.  There should be
a purely field theoretical argument which would prove the
existence of multiparticle excitations in regimes where the AdS
curvature is large, independently of the dimension of the space
or the existence of a weakly coupled string regime.  
Only when we understand this could we hope to check whether
a SUSY violating conformal field theory really represented
a large AdS space.

Finally, we would have to show that the theory contained metastable
excitations corresponding to black holes with size much bigger
than the Planck scale but much smaller than the AdS radius.  So far,
there is no evidence for nonsupersymmetric CFTs with these properties.
Indeed, we have little understanding of either cluster decomposition and
multiparticle structure , or metastable flat space black holes, in the
supersymmetric versions of AdS/CFT.

In summary, all the extant evidence indicates the absence of
asymptotically flat M-theory vacua with broken SUSY.  There 
are no solid examples, though the models of \cite{ks}
may yet turn out to fulfill their design criteria.
 
\section{The entropy of DeSitter space}

The results reviewed in the preceding section suggest (but
certainly not very strongly) that SUSY breaking in asymptotically
flat spacetime may be impossible in M-theory. There is certainly a
well known relation between the breaking of SUSY and the Ricci
scalar of spacetime.  Namely a generic nonsupersymmetric quantum
field theory generates a cosmological constant of order at least the SUSY
breaking scale. Conversely, a positive cosmological constant is
incompatible with SUSY.

The well known problem with this relation is the relative scale of
the two effects.  The cosmological constant is bounded from above
by a number of order eighty percent of the critical density, while the scale
of SUSY breaking is bounded from below by several hundred GeV.
Without fine tuning of parameters, and using the methods of
effective field theory, this leads to a cosmological constant
about 60 orders of magnitude larger than the observational bound.
We normally think about this problem by doing quantum field theory
in flat spacetime and then calculating the corrections to the
spacetime background.   SUSY breaking ``causes'' a large
cosmological constant which then makes the flat spacetime a bad
approximation.  I would like to suggest that we have been thinking
about this problem the wrong way around. The flat space
computation counts the zero point energy of the degrees of freedom
in spacetime.  We have been learning that the number and
properties of the degrees of freedom in M-theory depends crucially
on our specification of the boundary conditions on spacetime
\cite{bcdof}.  Asymptotically Anti-DeSitter spaces of various
dimensions have very different kinds of high energy degrees of
freedom and further they all differ drastically from
asymptotically flat spaces. Remarkably, the semiclassical
Bekenstein-Hawking formula consistently gives the right answer for
the extreme high energy entropy.  This is an example of the UV/IR
connection.  High energy states are associated with large, low
curvature (outside the horizon) geometries, whose gross properties
are encoded in general relativity.

For DS space, the Bekenstein-Hawking formula predicts a finite
entropy.  More precisely, any observer in an AsDS space only sees
a finite portion of the universe, bounded by a cosmological event
horizon.  One quarter of the area of this of this event horizon
(in Planck units) is the finite Bekenstein-Hawking entropy.  I
would like to interpret this number as the logarithm of the total
number of quantum states necessary to describe the universe as
seen by this observer.

There are three arguments for this.  The first is simply an
analogy with black hole physics (according to the holographic
principle ): event horizons may be viewed as holographic screens
on which all information about ``what is going on on the other
side of the horizon'' is encoded for the benefit of observers ``on
this side'' .  All of the arguments in favor of this holographic
view of black hole horizons apply equally well to DS space.

The second argument is by far the most convincing.  Imagine an
observer inside DS space trying to contradict our contention by
collecting as much entropy as she can. As long as she works on
scales smaller than the DS radius of curvature, she can do this
most efficiently by forming flat space black holes, whose entropy 
is bounded by their area. The black hole size is bounded by
something of order the horizon size so there is no way to violate
our bound.  Put another way, a system with an entropy larger than
the DS horizon size would simply not evolve into an AsDS spacetime
with the assumed value of the cosmological constant.

The third argument is more technical.  While few people believe
any longer that quantum gravity is described by an Euclidean
functional integral over metrics, this paradigm does seem to
provide helpful and correct hints about the quantum physics of
black holes and AdS spaces\cite{ghhpw}. Euclidean DS space is a
sphere, a compact geometry.  The rules for Euclidean quantum
gravity ({\it c.f.} perturbative world sheet physics in string
theory) tell us that all diffeomorphisms, including the DS group
of isometries are gauge transformations and should be integrated
over. All physical information is invariant. This is in marked
contrast to asymptotically flat or AdS universes, where the
isometries act nontrivially on the nonfluctuating boundary
geometry.  In these cases, the isometries are large gauge
transformations and physical states need not be invariant under
them.

Now consider quantum field theory in DS spacetime, defined by
analytic continuation of Euclidean Green's functions on the
sphere.  Long ago, constructive field theorists showed
\cite{chiara} for a large class of superenormalizable theories,
that these Green's functions have a Hilbert space interpretation
in terms of the Hilbert space of an observer living in the static
patch of Lorentzian DS space.  The state defined by these Green's
functions is the thermal state of the static patch Hamiltonian, at
the Hawking temperature.  These rigorous results are the
generalization of the observations of \cite{bd} for free field
theory and the perturbation expansion around it.  To obtain the
field theory in the full DS space one uses DS isometries to copy
the Green's functions from one static patch to another.  According
to the argument of the previous paragraph, this procedure just
produces gauge copies of the original system.  Thus, from this
point of view it would be wrong to introduce independent physical
degrees of freedom for each static patch.

It is important to examine several situations which appear to
contradict the idea that AsDS spaces have a finite number of
degrees of freedom.  One such argument is based on considering a
spacetime which is DS in the remote past.  At early times, the
volume of space is very large, and one can easily impose initial
conditions which have a larger entropy than the DS maximum.
However, most of these initial conditions will not lead to an AsDS
spacetime (with the same value of the cosmological constant).
Einstein's equations (with appropriate conditions on the stress
tensor) will not allow a violation of the Bekenstein-Fischler-Susskind
-Bousso (BFSB) bound\cite{bbmw}.

The ``approximately DeSitter'' spacetimes of inflationary
cosmology are confusing only so long as we forget the nature of
the holographic principle.  There is no cosmological event horizon
in these spacetimes (unless things settle down into a DS phase
much later in the history of the universe) , so the horizon size
of the inflationary DS phase is at best a temporary measure of the
maximal entropy in the experience of local observers.  When the
inflationary phase ends, the horizons of these observers expand.
The proper holographic screen on which all the information in
these universes can be encoded depends on their evolution after
the end of inflation.

By taking a limit in which the number of e-folds of inflation
becomes infinite, we can generate a paradoxical situation. If we
admit the possibility of independent information in different
static patches of DS space (as we have for any finite number of
e-foldings) then we obtain AsDS spacetimes with entropy larger
than the horizon area.  These are essentially the time reverse of
the spacetimes we encountered two paragraphs ago.  Of course, if
we extrapolate these expanding geometries back into the past, we
inevitably encounter a spacelike singularity.  Thus, the proper
description of these spacetimes is a Big Bang singularity which
evolve to DS space in the future.  Note that no local observer in
such a universe will ever encounter more entropy than is allowed
by the bound.  The confusion lies in the fact that there are many
ways of cutting the space up into regions observed by independent
local observers.  I believe that the confusion engendered by this
example is connected to initial conditions at the singularity, and
propose that a proper quantum treatment of cosmology will never
lead to spacetimes of this type.  In particular, I suspect that
general initial conditions at the singularity for a number
of degrees of freedom larger than the DS entropy will not
evolve into the postulated DS space.   The particular solutions
described above will involve extreme fine tuning of initial
conditions at the singularity, and might not exist at all in a 
quantum mechanical treatment.  

The claim that the cosmological constant determines the number of
degrees of freedom in an AsDS universe is extremely important if
true.  Traditionally, we think of the cosmological constant as an
effective field theory parameter with no direct connection to the
microscopic physics of the world.  It is to be calculated in terms
of more fundamental quantities. If however it is a direct count of
the number of degrees of freedom, then its value is part of the
fundamental set up of the quantum theory.  The dimension of
Hilbert space (if it is finite dimensional) or the number of
fundamental canonical degrees of freedom (if the Hilbert space is
infinite dimensional) is part of the definition of the theory. We
will see below that the possibility of such a direct connection
between an apparent low energy parameter and the fundamental
dynamics is an expression of the UV/IR relation of M-theory.

We must not attempt to calculate the cosmological constant but
rather to postulate its value and derive other observable
quantities from it.  From this point of view the ``cosmological
constant problem'' is turned on its head.  It is not ``why is the
cosmological constant so small'', but ``given the value of the
cosmological constant, why is SUSY breaking so large''.  Indeed,
although I cannot derive this logically from what I have already
said, in this context it seems inevitable that one should
attribute all breaking of SUSY to the fact that we live in an AsDS
universe.  This is consistent with the impossibility of defining
SUSY in DS space, and also with our failure so far to find SUSY
violating asymptotically flat states of M-theory, but it flies in
the face of all previous wisdom about SUSY breaking.

The classical formula relating SUSY breaking to the cosmological
constant is (without fine tuning) 
\eqn{susbrk}{M_{SUSY} \sim (\Lambda )^{1/4}.} 
A formula that fits the data is
\eqn{susbrkrt}{M_{SUSY} \sim (\Lambda M_P^4)^{\alpha}} with $\alpha =
1/8$.  I would propose that we describe these formulae with the
following slogan: {\it The $\Lambda / M_P^4 \rightarrow 0$ limit
of M-theory is a critical limit in which the number of degrees of
freedom of the system goes to infinity.  In this limit, the SUSY
breaking scale goes to zero, and we are trying to calculate the
critical exponent for its vanishing.  The classical mean field
value is $1/4$.  Experiment indicates that the correct value is
$1/8$.}

\section{How can this be?}

If the scale of SUSY breaking is smaller than the Planck scale,
then low energy physics is described by a locally SUSY effective
Lagrangian.  The breaking of SUSY in this Lagrangian is
spontaneous.  If the relevant SUSY is $N=1$ , $d=4$, then the
Lagrangian can have DeSitter solutions with spontaneously broken
SUSY.  The cosmological constant and the scale of SUSY breaking
are independent parameters in this Lagrangian.  The scalar
potential has the form

\eqn{efflag}{V = e^K[K^{i\bar{j}}F_i \bar{F_{\bar{j}}} - 3 \vert
W\vert^2 ]}

Everything has been expressed in Planck units.  We will be working
near the flat space limit, where the cosmological constant is very
small.  In that limit, the $F_i$ terms are the order parameters
for SUSY breaking in the sense that mass splittings in
supermultiplets are proportional to the values of the F terms at
the minimum of the potential.  Note that both {\it supermultiplet}
and {\it mass} are approximate concepts if the cosmological
constant is nonzero.  Mathematically, there are no global symmetry
generators with which to define these words precisely. Physically,
particles cannot be separated from each other by more than a
horizon size\footnote{See the section on the fate of the universe,
below.} , and we cannot define scattering amplitudes.

By choosing parameters in the superpotential and Kahler potential,
we can arrange a minimum with nonvanishing F terms and arbitrary
value for the cosmological constant.  However, this is generally
considered to be fine tuning, according to the following Wilsonian
argument.  When we calculate radiative corrections to the
effective Lagrangian below the SUSY breaking scale, we find a
contribution to the renormalized cosmological constant of order
$M_{SUSY}^4$, where $M_{SUSY}$ is the largest splitting in
supermultiplets, and is also chosen to be the cutoff in the
calculation.  This can be cancelled, by adroit choice of the
parameters in the Lagrangian, but the latter are thought to
represent the effect of integrating out fluctuations {\it at very
short spacetime scales}.  In local field theory, degrees of
freedom can be classified by their spacetime extent in an
underlying classical metric.  Degrees of freedom at short scales
see long wavelength degrees of freedom as essentially constant
parameters. According to this philosophy, the calculation of the
effects of short wavelength degrees of freedom is essentially
independent of the value of the cosmological constant, as long as
the latter is much smaller than (the - 4th power of) the
wavelength.  Thus, one argues, it is unnatural to imagine a
cancellation of the bare cosmological constant against the low
energy contribution.  Furthermore, in a field theory with
spontaneously broken SUSY, in flat spacetime, the very high energy
contributions to $\Lambda$ cancel.  Similar exact cancellations in
string theory with exact SUSY, suggest that this is not just a
fluke of the field theoretic approximation.

There are obvious problems with applying this argument to
M-theory.  The spacetime metric, which is used to characterize
what constitutes long and short wavelength fluctuations, is, in
M-theory, an approximate description of fluctuating quantum
variables.  More importantly, the association of large mass scales
with short distances is incorrect in M-theory. This
correspondence is valid down to the string scale in weakly coupled
string theory.  However, the high mass states of string theory are
predominantly of large spacetime extent.  More generally, above
the Planck scale, the high mass excitations are black holes, whose
Schwarzchild radius grows with their mass.  It is incorrect to say
that the dynamics of these objects is unaffected by the
cosmological constant.  Indeed, black holes with radius larger
than the cosmological horizon do not
exist in DS space.  Thus it is no longer implausible that the low
and high energy contributions to $\Lambda$ cancel each other
\footnote{An argument along these lines has been given by
L.Susskind in a variety of public and private venues over the last
six months.}.

Our identification of the cosmological constant as the (inverse
logarithm of) the number of quantum states of an AsDS universe
suggests a slightly different point of view.  The value of the
cosmological constant is now a fundamental parameter (actually a
boundary condition - see below) and we should set parameters in
our effective Lagrangian to match it.  In the low energy effective
Lagrangian, this requires us to find a vacuum with spontaneously
broken SUSY, but the natural scale of SUSY breaking is set by the
cosmological constant.  Field theoretic renormalizations will not
upset this relation.  There are, in Feynman diagrams, logarithmic
renormalizations of mass splittings in supermultiplets, but as
long as the field theoretic couplings are small, these are not
substantial when the cutoff is of order the Planck mass.
Furthermore, they do not depend strongly on the cosmological
constant.
Now however, consider quantum gravity corrections to the mass
splittings, first as loops of gravitons in Feynman diagrams. These
contribute to logarithmic divergences as well, but there is no
longer any small parameter controlling the series in powers of
logs.  However, there is still no apparent dependence on the
cosmological constant.  The crucial question now is what cuts off
the divergences when we reach the Planck scale.  Much has been
made of the softness of perturbative string amplitudes at large
momentum transfer \cite{gmetal} .  Many people have viewed this as
the ultimate cutoff promised by a true theory of quantum gravity.
But there is plenty of evidence, both internal to the perturbative
analysis\cite{veneziano}\cite{gmetal}\cite{tbf} and using D-brane 
techniques \cite{DKPS} that this is not correct.  In \cite{tbf} it was
suggested instead that the ultimate cutoff comes from black hole
physics.  That is, all high energy high momentum transfer
scattering amplitudes, and even the Regge regime, are eventually
dominated by black hole production with subsequent decay by
Hawking radiation.  This is again an invocation of the UV/IR
connection.  The gross features of the highest energy processes in
M-theory are ultimately encoded in General Relativity, because
they involve low curvature geometries.  We need the microscopic
theory to calculate the detailed quantum properties of the states
near a black hole horizon, but the level density of the high
energy spectrum and many properties of inclusive cross sections
can be calculated from semiclassical general relativity.

Thus, I would claim that there is no evidence for suppression of
"diagrams'' in which virtual black holes of mass much larger than
the Planck scale renormalize the splittings in low energy
supermultiplets.  The size of these contributions must be
estimated from the physics of black holes.  In such a calculation
it is clear that the DS horizon radius will provide a cutoff on
black hole contributions. It is entirely possible that a proper
calculation involves the detailed microphysics of black hole
states.  We will explore a more optimistic scenario below.

It is important to realize that there is no claim being made that
the theory with $\Lambda \rightarrow 0$ is divergent.  We are
merely trying to show that various quantities which vanish with
the cosmological constant do so more slowly than is indicated by
formulae which only take gravity into account classically. What we
{\it are} claiming is that the theory with vanishing cosmological
constant must be supersymmetric.  It is reasonable to suppose that
the restoration of SUSY will cancel otherwise divergent
contributions from virtual black holes.

\subsection{Proposal for a thermodynamic calculation}

Our proposal implies that a full understanding of the relation
between the cosmological constant and the scale of SUSY
breaking is possible only if we know something about M-theory
at very high energies.  Rather than giving up and saying
that this puts the problem beyond our powers at the present,
I would like to suggest that the UV/IR correspondence
may be used to get at least a rough estimate of the size of
the effect.  According to this principle, high energy physics
in M-theory is black hole physics, and some aspects of black
hole physics are computable in the semiclassical 
approximation to SUGRA.  We may hope that an estimate of
the relation between SUSY breaking and $\Lambda$ may be 
obtained in the semiclassical approximation.

The first aspect of semiclassical physics in DS space that
will be important to us is that the state of the system is
a thermal ensemble with respect to the static Hamiltonian
of DS space.  We consider this relevant, despite our previous
remarks that the DS group is a group of gauge transformations. We are
contemplating a limit of very small cosmological constant,
and trying to describe physics as seen by observers who
are unable to discern that space is not asymptotically flat
(because they are making observations that refer to low energy ,
approximately local, physics).
The phrase ``mass splittings in supermultiplets'' refers
precisely to properties of the (approximate) SuperPoincare
generators defined by such observers. 
The DS Hamiltonian goes over in the limit to the Poincare
Hamiltonian of the asymptotically flat observer.  We use it, because our
considerations will depend on the curvature of DS space.

Our second assumption is that the parameters in the local
effective Lagrangian actually get contributions from
``Feynman diagrams with virtual black holes in them''.
There is not even a semi-rigorous justification for this
assumption, and the following hand waving will have to suffice:
Consider Feynman diagrams contributing to the masses of
some of the particles in the theory.  As we allow the momenta 
in internal loops to grow larger than the Planck scale,
we encounter subgraphs which look like super-Planckian
scattering amplitudes, amplitudes in which all kinematical
invariants are larger than the Planck scale.  According
to classical general relativity, we expect such collisions
to result in black hole production.  I claim that the quantum
mechanical interpretation of this is that there is no
suppression of the probability of producing virtual black 
holes. 

The reader may be disturbed by the feeling that such large 
energy and momentum transfer processes should be cut off in
M-theory.  My response is that the black holes themselves
provide the cutoff.  For example, probability one black hole
production followed by Hawking evaporation, gives 
exponentially suppressed inclusive cross sections for finite
numbers of particles with energy and momentum transfer much
larger than the Planck scale\footnote{In weakly coupled 
string theory we find suppression of hard processes at a much
lower scale.  This however is only valid in an intermediate
regime \cite{tbf}.}. 

Given our two assumptions we expect the SUSY breaking mass
terms to be given by a thermodynamic average
\eqn{bhsum}{{\int dM e^{S(M) - \beta M} \Delta m(M) \over \int dM
e^{S(M) - \beta M}}}
Here $S(M)$ is the black hole entropy and $\beta$ is the
inverse Hawking temperature of DS space.  $\Delta m(M)$ is
the contribution to SUSY breaking from virtual black holes
of mass $M$.  We will restrict attention to four dimensions,
since this is the only place where low energy SUGRA can have
DS solutions.  In that case $S(M)= 4 \pi M^2 = \pi R_S^2
$, while $\beta = 2\pi R_D$.  The integral is actually cut off
when the Schwarzchild radius $R_S$ = $R_D$.  It is easy to
see that the integral is dominated by its upper endpoint
(unless $\Delta m$ falls extremely rapidly with black hole
mass).

Our claim then is the SUSY breaking
induced by DS space can be approximated by that due to
virtual black holes of a size near the upper cutoff for
Schwarzchild-DeSitter black holes.  I hope to report on
an estimate of this effect in the near future.

\section{The fate of observers in an AsDs universe}

There is a line in an old country and Western song that goes
"DeSitter space is a lonely place $\ldots $'' .  Indeed, once
the cosmological constant takes over the expansion rate,
everything that is not gravitationally bound to us soon passes
outside our horizon.  Worse, after baryons decay, gravitationally bound
systems will cease to exist if they have not collapsed into
black holes.  And when quantum mechanics is taken into account
even this ultimate refuge is lost to us, since the black holes
decay.  Eventually, the universe becomes full of elementary
systems, each in its ground state in its own horizon volume
(we are for the moment ignoring the Hawking radiation of
DeSitter space).

Physics as we know it, which describes local interactions
between systems which can communicate with each other, becomes 
increasingly irrelevant in such a universe, though the
time scale for this to happen is enormously long.  Thus, the
usual apparatus of physics describes an epiphenomenon in
an AsDS universe.  One of the technical problems related to
this observation is how one describes the physical answers
that are relevant to us as exact, gauge invariant, mathematical
quantities in such a theory.   

In asymptotically flat space, the holographic principle tells
us that we can calculate the S-matrix.  So far we have 
found no other sensible physical quantities in Asymptotically 
Flat M-theory.   But there is no S-matrix in AsDS spaces.
One must really search for more local quantities, but it
seems that any such search may have only an approximate nature.
For example, one might imagine showing that the 
low energy effective Lagrangian description had the status
of the first term in an asymptotic expansion of something.  
But what might that something be?  If we extrapolate to high
enough energy we are always required to ask questions 
about all of the degrees of freedom and their dependence on
the global geometry of AsDS space.  There is no exact quantum
number that takes the place of energy.   
If we are willing to take the attitude that at sufficiently
high energy we can neglect SUSY breaking, we can use the
flat space, SUSY vacuum which best approximates our AsDS
universe to calculate scattering amplitudes above the Planck
scale.  But we must recognize that at sufficiently high 
energies these amplitudes describe processes involving black
holes larger than the DS radius.  These have nothing to
do with anything in the real world, if the universe is AsDS.
It is also far from obvious to me that one could find 
a systematic incorporation of the the SUSY violating corrections to
these amplitudes into a more exact description of the
world.  In our view, SUSY violation is a consequence of
the AsDS geometry of the universe, and might be incompatible
with a description of the world in terms of scattering 
amplitudes.  The phrase {\it SUSY violating scattering amplitude}, might
be an oxymoron that made sense only at energies below the
Planck scale.

All of this suggests that there is a somewhat more local
description of holographic physics than any which exists
at present.  I presented a preliminary sketch of what such
a formalism might look like at the Millenium conference
in January.  It involves a collection of Hilbert spaces
${\cal H}_i$, each of which is supposed to represent
those states observable in the causal past of a finite number of points,
in a cosmological spacetime which begins at a Big Bang singularity.  
More precisely, using the
Bekenstein-Hawking-Bousso relation between areas and 
entropy, and a causal structure which is defined by 
mappings of the algebra of operators in one space into
a subalgebra with (in general) 
nontrivial commutant in another, I proposed to reconstruct
a spacetime directly from quantum mechanics.
\footnote{This is not the place for a detailed exposition
of these ideas, which I hope to present at a later time.\cite{tbfut}}

In this formalism, the experience of a more or less localized
observer is encoded in a sequence of Hilbert spaces of
(exponentially) increasing dimension.  Each space in the
sequence is mapped into a tensor factor of the one succeeding
it.  In order to have unitary evolution, the full state in
the successor Hilbert space must be determined by partial
mappings from many different predecessor states.  In general
it is not required that the entire process, including an
infinite sequence of steps, can be incoporated
in a single Hilbert space of finite dimension. One consistent
rule which allows this is that the Hilbert spaces in any sequence
converge after a finite number of steps to a space of some
fixed dimension, the same for every sequence.  The inclusion
maps become unitary mappings of this space into itself.

I would like to identify such a situation with an AsDS space,
in the limit that the number of dimensions of the asymptotic
Hilbert space is very large.  Appropriately smooth unitary
mappings between different sequences would represent the
different ways in which the spacetime could be represented as
the static patch of a given observer, each of whom perceives 
all of the things outside her horizon as a thermal gas.

In this view of the universe, the local degrees of freedom
whose investigation is the province of experimental physics
should be viewed as being "on temporary loan'' from the
"thermal DeSitter library'' .   As the DeSitter era unfolds  
, more and more of the observer's degrees of freedom are
"returned to the shelf'': they get swept outside his horizon,
and become part of the thermal background.  It is interesting
that the total number of borrowed degrees of freedom that
we need to describe what we see is, even if we include
the entropy in hypothetical black holes in the center of each
galaxy, smaller by a factor of $10^{30}$ than the Bekenstein
Hawking entropy corresponding to the cosmological constant.
Thus, from a sufficiently cosmic viewpoint, the entire 
organized part of
the universe may be just a small coherent fluctuation
in a random system with an enormous number 
(nearly a googleplexus)of degrees of freedom.
It may be that in the far future, after the universe has degenerated
into a collection of frozen elementary systems, each in its
own horizon volume, a new fluctuation in the Hawking radiation
can form, and the whole process will begin again.

Let me conclude this section by repeating that the most 
important technical problem posed by this view of the AsDS 
universe is to realize the physical measurements we make
in terms of exact mathematical statements about the finite
dimensional Hilbert space associated with the spacetime.

\section{Metaphysics}

One of the most disturbing aspects of the proposal in this paper
is that the theory of the universe involves a fixed integer $N$,
the total number of quantum states in the universe. I believe that
a discussion of the meaning of this number will depend on the
distinction between equations of motion and boundary conditionsin
physics.  
It has long been apparent, that even if we find the
ultimate physical laws encoded in a set of equations of motion, we
will still have to deal with the question of what determines the
boundary conditions.  In cosmology, this question has
traditionally been split into two parts: "Do the spatial sections
of a Friedmann-Robertson-Walker cosmology have a boundary (and
what are the boundary conditions there)?'', and "What are the
initial conditions?".   Einstein preferred closed cosmologies
because he believed this eliminated the first of these questions.
Various authors \cite{HHetal} have tried to address the second.

There is a well known problem associated with Einstein's
suggestion, if one believes that quantum theory is the ultimate
description of nature, and also believes in an ultraviolet cutoff.
A closed universe with a UV cutoff must have a finite number of
states.  If we try to associate the cutoff with a cutoff of short
distances, we immediately run into a problem.  The volume of the
universe changes with time, so the number of states allowed by a
short distance cutoff would appear to change as well.  This
violates unitarity.

The advent of holographic cosmology \cite{fsb}\cite{bbmw} has resolved this
conundrum.  The obvious conjecture that follows from this work is
that the number of states in a cosmology is the exponential of one
fourth of the area in Planck units of a maximal set of holographic
screens\footnote{Provision must probably be made for duplication
of information on different screens.  An incomplete draft of a 
proposal for a completely
quantum mechanical and holographic formalism for cosmology was
outlined in my talk at Strings at the Millenium in CalTech.  A
somewhat more extended presentation of these ideas is in
preparation\cite{tbfut} .}.  I believe that ultimately this
prescription will be turned around.  Cosmology will be derived
from quantum mechanics, with spacetime geometry being computed
from the number of quantum states.

From this point of view, the natural distinction between
cosmological boundary conditions will be in terms of the number of
quantum states that they admit.  We first have the possibility of
a finite number, and then infinity.  We expect that systems with a
finite number of states can describe either AsDS universes or
recollapsing universes.  It is likely that the distinction between the
two is simply whether we require an infinite or a finite number
of steps in our choice of time evolution.

 With an infinite number of states it is
natural to look for some operator on the Hilbert space whose
eigenspaces with finite eigenvalue are finite dimensional and then
to make a finer classification in terms of the behavior of the
density of states at large eigenvalue. Geometrically we would
expect this to map into the problem of black hole entropy in
cosmologies with no finite area cosmological horizon.

I think that, apart from the apparent observational evidence for a
cosmological constant, our reaction to the choice between finite
and infinite cosmologies (in the present sense) can at best be an
emotional one.  On the one hand, it is reasonable to think that
nothing is actually infinite - that infinity or infinitesimal
always refers to an idealization that makes problems more easy to
treat mathematically (in the practical, rather than the rigorous
sense of mathematics).  Then one will be saddled with the annoying
question of why a particular finite number is chosen. This may
lead one prefer to accept infinity as a reality, though I would
claim that the various choices among behaviors of the asymptotic
spectrum of black holes will be equally annoying.  One may find
that insisting on a large asymptotic symmetry group somewhat
restricts the possibilities, but the plethora of exactly stable
Poincare and AdS vacua of M-theory makes this seem unlikely. The
only theoretical basis for resolving this problem would seem to be
to prove a theorem that every system with an infinite number of
states which is asymptotically describable by a large smooth
geometry, becomes supersymmetric in the asymptotic limit.  As, I
have noted, there is some meager evidence for this conjecture.

Given that N is finite, the question of how it is chosen might
have two generic kinds of answer:
\begin{itemize}
\item In the fullness of time we might show that N had to satisfy
some number theoretic property that is satisfied by $[0,1,2, 216,
2^{{10}^{120}} +23 , 2^{{10}^{250}} + 13365 , \ldots ]$ .  Or
perhaps it is the unique solution to some number theory problem.

\item There is some meta-dynamics which gives rise to quantum
systems with different values of N\cite{hm}. 
Perhaps it is even some 
kind of deterministic dynamics and could alleviate our
unease with the application of probabilistic ideas to the 
whole universe.  In such a system we might find either a true dynamical
explanation of the value of N, or the framework for an anthropic
determination of this single parameter.

\end{itemize}

The point about these possible answers is that they have very
little to do with physics in the universe we observe (hence the
title of this section). Our best strategy is probably to ignore
the question. The most useful attitude would appear to be to
assume N is a boundary condition and hope that many features of
the dynamics have universal properties for large but finite N.
Thus the characterization of the formula $M_{SUSY} \sim
\Lambda^{1/8}$ as a formula for a critical exponent.

\section{Some remarks on phenomenology}

One of the most interesting features of the proposal in this paper
is that it solves what I consider one of the primary
phenomenological problems of M-theory, namely why we do not live
in one of the many stable supersymmetric ground states of the
theory.  The answer is simply that we do not have enough states.
Poincare invariant ground states have an infinite number of
excitations, at least all of the scattering states of gravitons.

 Our suggestion about
the origin of SUSY breaking probably has more practical
implications for SUSY phenomenology as well. For example suppose
that the generation structure of the standard model is related to
a discrete gauge symmetry that is spontaneously broken at an
energy scale well below the Planck scale.  We have attributed the
dominant contribution to SUSY breaking to very high energy black
hole states.  These states will be insensitive to the low energy
breaking of generation symmetry and might well produce flavor
singlet squark mass matrices. Alternatively, the mere fact that
SUSY breaking comes from a thermal average over a large number of
states might produce flavor singlet mass matrices, without appeal
to symmetries (Of course, we probably want to have flavor
symmetries to explain the quark mass matrix.).
 One might imagine the
possibility of deriving the minimal SUGRA spectrum, or some other
simple pattern of SUSY breaking, from this scenario.

Another general conclusion would appear to be 
that the gravitino mass, as well as the
masses of any moduli which originate from SUSY breaking, will be
of order $\Lambda^{1/4}$.  This causes well known cosmological
difficulties, which must be solved.

Finally one may hope that the current approach to
cosmology will eventually solve the vacuum selection
problem of string/M-theory.  In the limit of vanishing cosmological
constant, our approach implies that 
the finite dimensional Hilbert space of an AsDS M-theoretic cosmology,
approaches that of an asymptotically flat SUSY vacuum of
M-theory.  This is presumably the state which describes
scattering of particles inside gravitationally bound clusters during the
pre-asymptotic stage of the AsDS universe.

The question of which flat SUSY background we approach in the limit {\it
might} depend on initial conditions - that is, in the small $\Lambda$
limit, the Hilbert space might break up into superselection sectors and
different cosmological evolutions might end up in different sectors.
On the other hand, one might hope for a more unique and universal
answer. At any rate, the question is certainly tied up
with that of initial conditions for cosmology.

Certain features of the desired background can be 
understood from general considerations.  It must be supersymmetric, and
its low energy effective Lagrangian must have 
a small deformation corresponding to a SUSY violating DS space.
This makes it virtually certain that the SUSY background
cannot have any moduli.  Small deformations of
a SUSY Lagrangian with moduli will generally give rise
to cosmologies with varying moduli, rather than a
DS space.   In  \cite{tbcosmo} I discussed a general analysis of
inflationary cosmologies
deriving from M-theory.  Approximate moduli were argued
to be good inflaton candidates, and the discrepancy between the
inflation and SUSY breaking scales was attributed to 
the existence to a submanifold of approximate moduli space where SUSY
and a discrete R symmetry were restored.  Much of the postinflationary
dynamics of the universe depended on the dimension of this
submanifold.  The present considerations suggest that
one wants it to be a point, as has long been advocated by Dine \cite{dine}.
This suggests that, in order to find the vacuum state of M-theory that
describes the universe approximately,  one must search for 
an isolated point in the approximate moduli
space of an $N=1$ compactification, which preserves SUSY
and a discrete R-symmetry.

\section{Conclusions}
 
It should be obvious that the claims made here are somewhat
tentative and unformed.  One aspect of the subject that I find 
rather confusing is the relation of the fundamental theory
to the low energy effective Lagrangian.  Despite the UV/IR
correspondence, I believe it is correct that physics below
the Planck scale is governed by a locally supersymmetric
effective Lagrangian.  In \cite{tbfut} I have suggested that
local SUSY is in fact connected to the arbitrary choice
of holographic screen, and should therefore be a fundamental
symmetry, not to be broken.  Since we expect the scale of
SUSY breaking to be much smaller than the Planck scale
there should be an effective Lagrangian description of low energy
physics which is locally supersymmetric, which means that SUSY breaking 
appears spontaneously.  The SUSY breaking scale and cosmological 
constant should simply be set by tuning parameters in this Lagrangian.

The confusing point is that in this description there appears to be a
low energy origin for SUSY breaking.  Some chiral field's F term gets
a nonzero expectation value.   I suspect that the correct
description will simply introduce SUSY breaking through
a Volkov-Akulov \cite{va} goldstino multiplet.  The SUSY breaking scale
and cosmological constant will be put in by hand.  They are related by a
formula of the form $M_{SUSY} = K M_P (\Lambda / M_P^4)^{1/8}$.  This
formula can only be understood, and the constant $K$ 
calculated, within the framework of the full theory.
Similarly, the couplings of the Goldstino to other
low energy fields, which determine the phenomenology of SUSY breaking,
will depend on high energy physics.  Only if the conjectures about
relating high energy physics to black hole physics, which were
adumbrated in section 4 are correct, will we be able to
extract any details of the SUSY spectrum without a full understanding of
the quantum mechanics of M-theory.
\acknowledgments

I would like to thank Willy Fischler for impressing upon me the
importance of understanding the entropy of DeSitter space and for
numerous conversations about this subject.  Raphael Bousso, Lenny
Susskind, and Emil Martinec also made very useful comments and suggestions.
This work was supported in
part by the Department of Energy under grant DE-FG02-96ER40559.
%
 

\end{document}